\documentclass{aip-cp}

\usepackage[numbers]{natbib}
\usepackage{rotating}
\usepackage{graphicx}

\begin{document}

% Title portion
\title{Highlights from the VERITAS AGN Observation Program}

\author[aff1]{W. Benbow\corref{cor1}}
\affil[aff1]{Harvard-Smithsonian Center for Astrophysics, 60 Garden
  St, Cambridge, MA, 02180, USA}
\author[aff2]{the VERITAS Collaboration}
\affil[aff2]{http://veritas.sao.arizona.edu/}
\corresp[cor1]{Corresponding author: wbenbow@cfa.harvard.edu}

\maketitle

\begin{abstract}
The VERITAS array of four 12-m imaging atmospheric-Cherenkov
telescopes began full-scale operations in 2007, and is one of the 
world's most sensitive detectors of astrophysical very high energy
(VHE; E$>$100 GeV) gamma rays. Observations of active galactic 
nuclei (AGN) are a major focus of the VERITAS Collaboration, and 
more than 60 AGN, primarily blazars, are known to emit VHE photons.  
Approximately 4000 hours have been devoted to the VERITAS AGN 
observation program, resulting in 34 detections. Most of these
detections are accompanied by contemporaneous, broadband 
observations, enabling a more detailed study of the underlying 
jet-powered processes. Recent highlights of the VERITAS AGN 
observation program are presented.
\end{abstract}

\section{INTRODUCTION}
AGN are among the most powerful particle accelerators in the universe,
and emit non-thermal radiation from radio through gamma rays. 
As of September 2016, 63 AGN are identified as VHE sources, and
these objects comprise about one-third of the VHE sky catalog\cite{TEVCAT}.
They are the most numerous class of identified VHE $\gamma$-ray
emitter, and those detected in the VHE band all belong to the small 
fraction of AGN possessing jets powered by accretion onto a 
supermassive black hole (SMBH).  It is generally believed that the production of
VHE $\gamma$-rays occurs in a compact region within these jets, near
the central SMBH.

Blazars are a class of AGN with jets pointed along
the line-of-sight to the observer, and most ($\sim$94\%) of the 
VHE $\gamma$-ray emitting AGN are blazars. 
The VHE blazar population includes
four blazar subclasses:  47 high-frequency-peaked BL\,Lac objects (HBLs),
8 intermediate-frequency-peaked BL Lac objects (IBLs), 
2 low-frequency-peaked BL Lac object (LBL),
and 5 flat-spectrum radio quasars (FSRQs), as well as one
blazar whose sub-classification is uncertain. 
VHE blazars are detected over a redshift range of $z = 0.030$
to $z=0.944$, but most of these objects have redshift $z < 0.3$.
Energetics requirements and the attenuation of VHE photons on the extragalactic background
light (EBL), are the major contributors to the observed redshift distribution.
Four nearby ($z<0.06$) FR-I radio galaxies are also detected at VHE.

VHE AGN are generally variable.  About one-third are only
detected at VHE during flaring episodes.   The remaining AGN generally
show VHE flux variations of a factor of 2-3 on timescales ranging from
days to years, similar to their behavior at other wavelengths.
Although there have been several noteworthy events (see, e.g.,
\cite{PKS2155_flare}), the detection of rapid
(minute-scale), large-scale (factor of 100) variations of the VHE flux 
remains relatively rare.   The VHE photon spectra observed from AGN are often soft 
($\Gamma_{obs} \sim 3 - 5$), and rarely is any emission observed
above $\sim$1 TeV.  This is partly because the VHE band is generally
located above the high-energy peak of the AGN's double-humped 
spectral energy distribution (SED), and also because of softening of 
the emitted AGN spectra via EBL absorption.

The VERITAS collaboration hopes to improve the understanding of VHE
AGN and their related science by making precision measurements of
their spectra and their variability patterns.  The VERITAS AGN program is largely focused on long-term studies of the
existing VHE AGN population and the search for, and the observation of, major flaring episodes.
All VERITAS AGN studies are accompanied by contemporaneous multi-wavelength (MWL) 
observations to enable modeling of the AGN SEDs, as well as searches for
correlations in the observed flux / spectral changes that may indicate
commonalities in the origin of the AGN emission.  VERITAS AGN studies
can also be used to make cosmological measurements such as
the density of the EBL and the strength of the intergalactic magnetic field (IGMF).
More details on the VERITAS EBL / IGMF effort can be found in these proceedings \cite{Elisa_HDGS2016}.

\section{VERITAS AGN Program}
VERITAS began routine scientific observations with the full array in 
September 2007, and was significantly upgraded in Summer 2012,  improving its low-energy
response.  The observatory is most sensitive between $\sim$85 GeV and $\sim$30 TeV and 
can detect an object with flux equal to 1\% Crab Nebula flux (1\% Crab) in $\sim$25 hours. 
Spectral reconstruction of observed signals can be performed above
$\sim$100 GeV, and typical systematic errors are $\sim$20\% on the
flux and 0.1 on the photon index ($\Gamma$).

From 2007 to 2016, VERITAS acquired an average of $\sim$950 h of good-weather observations each year
during ``dark time''.   Of these observations, a total of $\sim$3800 h
were pointed at AGN ($\sim$425 h per year) with $\sim$90\% split to blazars,
primarily BL Lac objects, and $\sim$10\% to radio-galaxies, primarily M\,87.  
Beginning in September 2012, the VERITAS
collaboration developed the capability to observe during periods of
``bright'' moonlight (i.e. $>$30\% illumination), adding another 
$\sim$30\% to overall good-weather data yield. These observations have
only slightly higher threshold (e.g. 250 GeV) and have been used to successfully
detect a flaring blazar \cite{VERITAS_1727}.  Since 2012,
approximately 700 h of good-weather, bright-moon
observations ($\sim$175 h per  year) were targeted at AGN.  Overall,
nearly half the acquired VERITAS data are AGN observations, and AGN comprise 61\% of the VERITAS source
catalog.  A table listing the 34 AGN detected by VERITAS can be found
in \cite{Benbow_ICRC2015}.  

The philosophy of the VERITAS AGN (radio galaxy and blazar) program has shifted from an
emphasis on expanding the source catalog by discovering new VHE AGN,
to exploiting the existing catalog via deep / timely measurements of
the known sources.  Only 20\% of the VERITAS blazar program is currently devoted to the discovery and
follow-up observations new VHE AGN (c.f. $\sim$80\% from 2007-2010).
Similarly the radio-galaxy program was initially $\sim$40\%
VHE discovery efforts, and only 3 h of discovery data have been taken the past
four seasons.  The VERITAS AGN program is now
heavily devoted to regular monitoring of the entire Northern VHE catalog,
with a cadence designed to generate deep exposures for some
particularly interesting targets. These monitoring observations are
coordinated with partners at lower energy, so that long-term
contemporaneous MWL data sets exist for all Northern VHE AGN.
Should any interesting flaring events be observed in these
monitoring programs, intense MWL target-of-opportunity (ToO)
observations are planned.   ToO observations are a key component of the VERITAS AGN 
program, averaging  $\sim$30\% of these data each year.  Nearly all VERITAS FSRQ data are taken via ToO observations.

\section{Recent Highlights}

\begin{figure}[]
   \centerline{ {\includegraphics[width=3.0in]{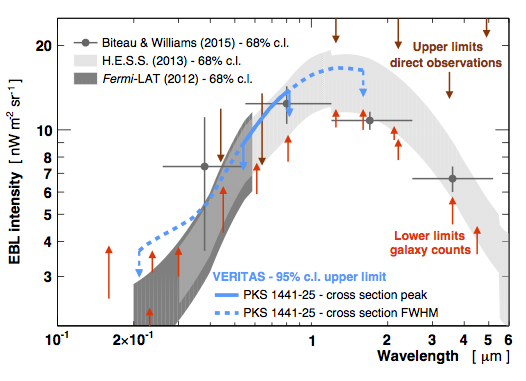} }
              \hfil
              {\includegraphics[width=3.0in]{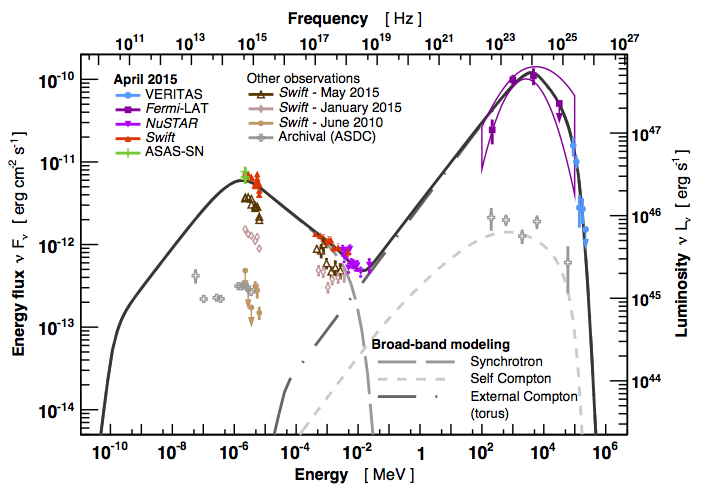} }
             }
  \caption{Left: VERITAS
    constraints on the EBL SED derived from PKS 1441+25. Right: SED of PKS 1441.  }
\label{PKS1441_plot}
\end{figure}

{\bf PKS\,1441+25} is an FSRQ located at a redshift of $z = 0.939$.
It was discovered as a VHE $\gamma$-ray emitter by MAGIC \cite{MAGIC_1441} 
in April 2015, following a flaring alert from  {\it Fermi}-LAT.
These flare detections triggered a week-long ToO observation
campaign with VERITAS, resulting in the detection ($\sim$8$\sigma$) of a very soft spectrum
($\Gamma = 5.3 \pm 0.5$) excess of $\sim$400 events in 15 h of good-quality observations
\cite{VERITAS_1441}.  The observed flux was steady at $\sim$5\% Crab
above 80 GeV during the ToO period, which also coincided with a period of enhanced MWL emission.
No significant excess was observed from the source during a further $\sim$4 h of data taken in May 2015,
after the MeV-GeV ({\it Fermi}-LAT) flare had subsided, nor was any
seen during snapshot monitoring in 2016. After correcting the observed {\it Fermi}-LAT and VERITAS spectra for
EBL effects, the two data sets connect smoothly.  This suggests that it
is not unusual that VERITAS detected this object below 200 GeV, despite its
redshift of $\sim$1.  Indeed, the VERITAS detection yields limits on
the EBL that are comparable to the strongest known (see
Figure~\ref{PKS1441_plot}), and are consistent with recent models.  Of particular
note in the VERITAS study is the copious amount of MWL data.
Highlights include long-term correlations, with no delay, in the radio, optical and {\it Fermi}-LAT light
curves suggesting a single-emission region is responsible for the
emission, and X-ray data showing synchrotron emission up to $\sim$30
keV, along with a variability time scale of less than 2 weeks. A
synchrotron self-Compton (SSC) plus external Compton (EC) model was
fit to the contemporaneous data (see Figure~\ref{PKS1441_plot}), and it yields normal parameters
(e.g., a low doppler factor, a system near equipartition) except for a
noticeably large emission region ($\sim$200,000 Schwarzchild radii).
Coupling the conclusions from the MWL observations together with
opacity arguments that require the gamma-ray emission region be
outside the quasar broad-line region, provides the first strong
evidence for large-scale VHE emission from an AGN.

On February 8, 2014, an exceptionally bright flare of {\bf B2\,1215+30} was observed during routine
VERITAS monitoring of 1ES\,1218+304.  The VHE flux reached 240\% Crab, more than
60 times the lowest value previously recorded from this HBL.  This flare lasted less
than a single day, and the observed variability time scale is less
than 3.6 h.  The measured flux corresponds to a rarely-seen isotropic luminosity
of $\sim$$2\times10^{46}$ erg s$^{-1}$.  For comparison, Mrk\,421
would need to be 35 times brighter than the Crab Nebula to reach a
similar luminosity.  The VERITAS flare was correlated with a high GeV
state seen by  {\it Fermi}-LAT, but no optical counterpart was
observed by the Tuorla observatory.  More details on this event and
the interpretation of these data within an SSC framework can be found
in \cite{B21215_paper}.

\begin{figure}[]
   \centerline{ {\includegraphics[width=5.0in]{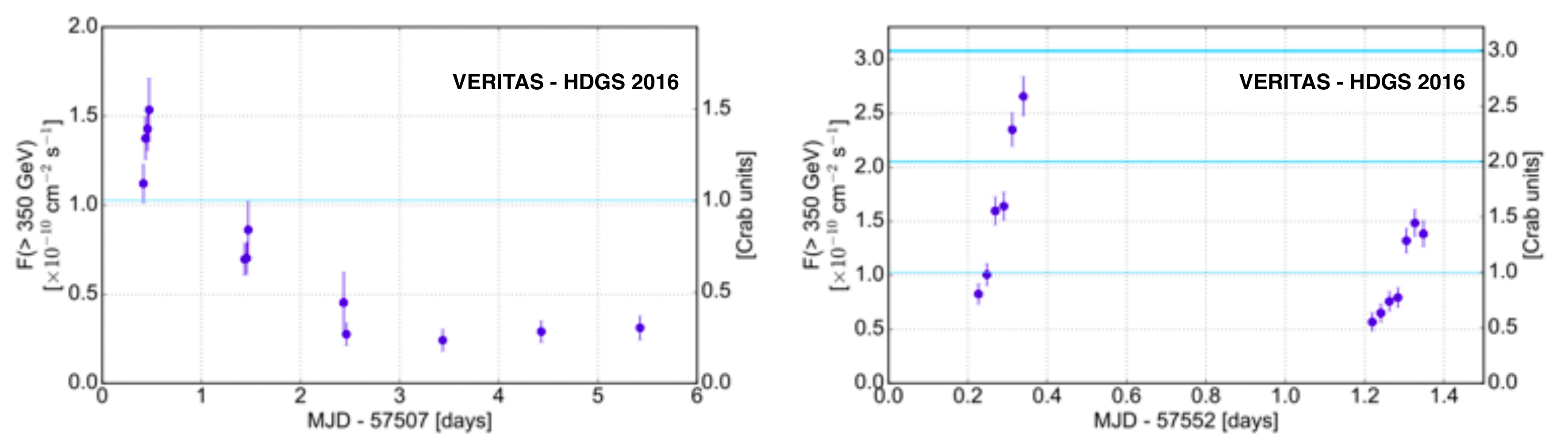} }
             }
  \caption{Light curves  for 1ES\,1959+650 in April 2016 (Left) and June
    2016 (Right).}
\label{1ES1959_plot}
\end{figure}

{\bf 1ES\,1959+650} is a well-known VHE emitter, whose possible "orphan flare" in
2002 prompted significant scientific interest.  It has been relatively
inactive since this event, but beginning in July 2015, an
extended elevated state was observed during multi-wavelength
observations of this nearby ($z = 0.047$) HBL.  The X-ray and MeV-GeV
fluxes were the highest seen from this source since the start of the
{\it Swift} and {\it Fermi}-LAT missions.  VHE flaring was also observed by
VERITAS in October-November 2015, peaking at $\sim$1 Crab (ATel
\#8148).  The MWL high state persisted into 2016, and 1ES 1959+650 was
observed again by VERITAS for 13 nights between April and June 2016.  
During this time more than 2400 photons were detected ($>$80$\sigma$)
in $\sim$12 h of quality-selected data.  Multiple $>$1 Crab flares
were observed (see Figure~\ref{1ES1959_plot}), including events in April (ATel \#9010) and June (ATel
\#9148).  The VHE flux peaked at $\sim$2.5 Crab on
MJD 57552 (June 13).  The preliminary spectrum generated from these
data extends to $\sim$8 TeV and has a photon index of $\Gamma = 2.6$.

\begin{figure}[]
   \centerline{ {\includegraphics[width=3.5in]{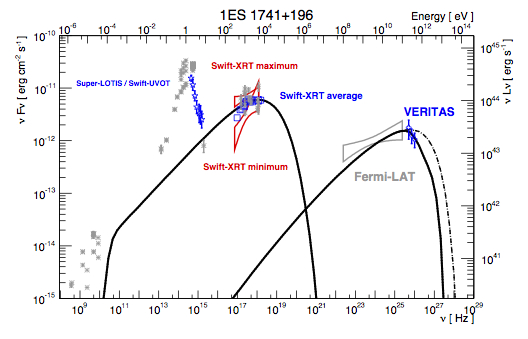} }
              \hfil
              {\includegraphics[width=2.25in]{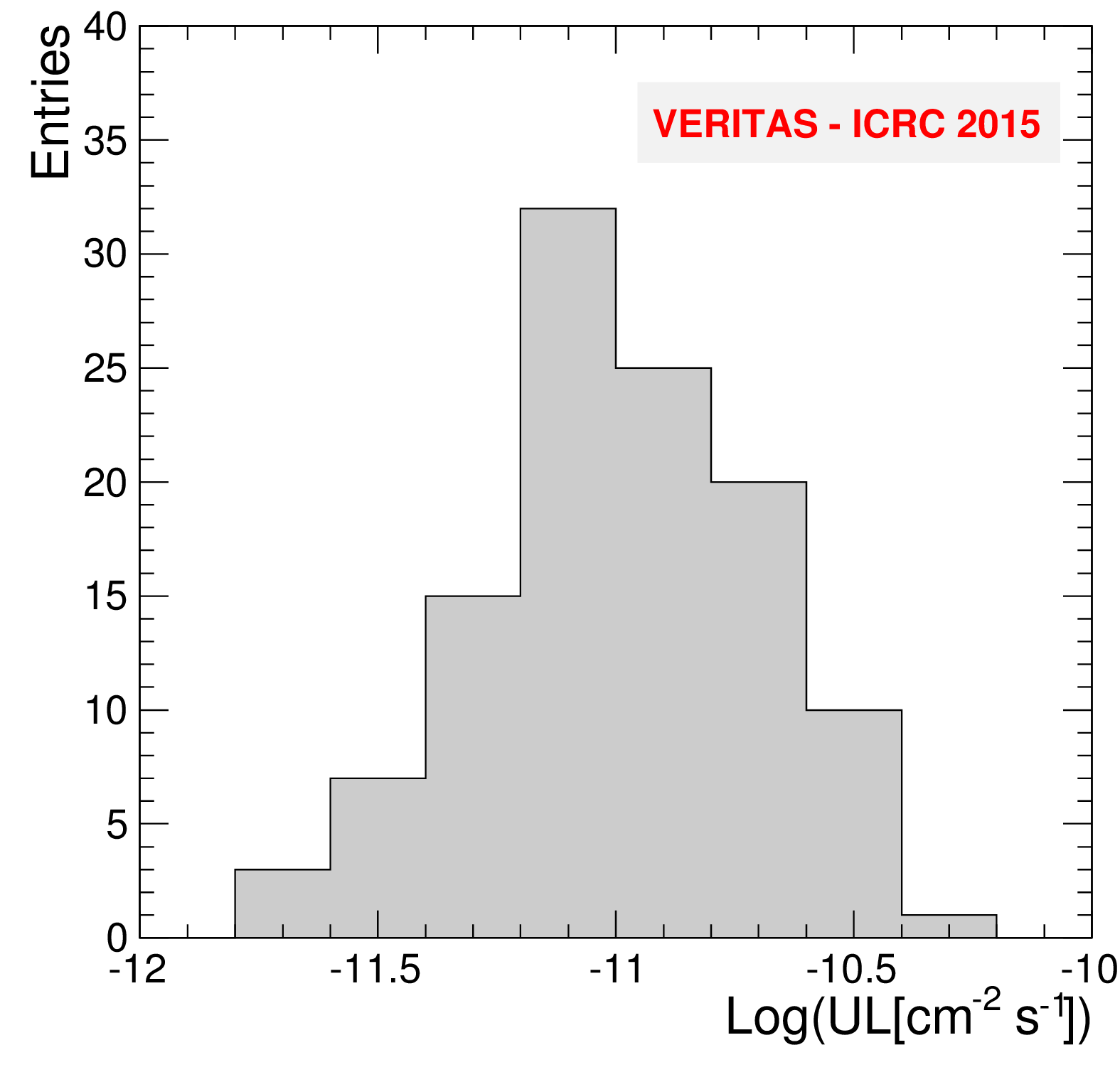} }
             }
  \caption{Left:  SED of 1ES\,1741+196.  Right:  Histogram of integral flux upper limits derived from
     VERITAS observations of 93 blazars and all 21 other 2FGL sources in
     the field of view of those blazars.}
\label{AGN_mix_plot1}
\end{figure}

{\bf 1ES\,1741+196} is a nearby ($z = 0.084$) {\it Fermi}-LAT-detected HBL.  It was identified as a likely VHE emitter
on the basis of its SED (see, e.g., \cite{Luigi_AGN}) and its MeV-GeV
spectrum ($\Gamma_{3FGL} \sim1.8$), and was
discovered as a VHE emitter following extensive observations by MAGIC
in 2011.  Its initial detection placed it among the weakest of the known VHE
sources.  VERITAS observed 1ES\,1741+196
for $\sim$30 h of quality-selected live time from 2009 to 2014.  In
these data an excess of 120 $\gamma$-rays
($\sim$5.9$\sigma$) was detected from the direction of the blazar \cite{VERITAS_1741}.
The observed flux of $\sim$1.6\% Crab above 180 GeV is marginally higher than, but consistent with,  the
MAGIC flux ($\sim$0.8\% Crab).  The VERITAS spectrum is relatively hard
($\Gamma = 2.7 \pm 0.7$) for a blazar, which is perhaps due to this object's low
redshift and relatively high inverse-Compton peak frequency.
The SED (see Figure~\ref{AGN_mix_plot1}) generated from contemporaneous VERITAS,  {\it Fermi}-LAT,
{\it Swift} XRT/UVOT and Super-LOTIS data is compatible with an SSC model.

{\bf HESS\,J1943+213} is a point-like, $>$500 GeV emitter found in the
HESS Galactic Plane scan \cite{HESSJ1943}. It is a hard-spectrum
{\it Fermi}-LAT source and might be a blazar, but no flux variations have been observed, nor has any
redshift been measured.  VERITAS observed HESS\,J1943+213 for 31 h
of good-quality live time in 2014-2015.  A strong excess (20$\sigma$;  3.6$\sigma$ h$^{-0.5}$) was
observed, corresponding to a flux that is consistent with the HESS (2005-08)
measurement.  No flux or spectral variations are seen on any time
scale, and the observed flux ($\sim$2.2\% Crab) and photon index ($\Gamma = 2.8\pm0.1$)
are consistent with the HESS results.  More details on these VERITAS
studies, along with recent VLBA observations of HESS\,J1943+213, can be found in \cite{Karlen_HDGS2016}.

\begin{table}[]
\begin{tabular}{c | c | c | c |c}
\hline
{\footnotesize AGN} & {\footnotesize $z$} &  {\footnotesize Type} &
{\footnotesize log$_{10}(\nu_{\rm synch})$ [Hz]} & {\footnotesize
  {\it Fermi}-LAT Name}\\
\hline
{\footnotesize W\,Comae} & {\footnotesize 0.102} & {\footnotesize IBL}
& {\footnotesize 14.8$^{\dagger}$} & {\footnotesize 3FGL J1221.4+2814}\\
{\footnotesize RGB\,J0521.8+2112} & {\footnotesize 0.108} &
{\footnotesize IBL} & {\footnotesize 15.1} & {\footnotesize 3FGL J0521.7+2113}\\
{\footnotesize RGB\,J0710+591} & {\footnotesize 0.125} & {\footnotesize HBL} & {\footnotesize 18.1} & {\footnotesize 3FGL J0710.3+5908}\\
{\footnotesize S3\,1227+25} & {\footnotesize 0.135} & {\footnotesize
  IBL} & {\footnotesize 15.0} & {\footnotesize 3FGL J1230.3+2519}\\
{\footnotesize 1ES\,0806+524} & {\footnotesize 0.138} & {\footnotesize HBL} & {\footnotesize 15.9} & {\footnotesize 3FGL J0809.8+5218}\\
{\footnotesize 1ES\,1440+122} & {\footnotesize 0.162} & {\footnotesize HBL} & {\footnotesize 17.2} & {\footnotesize 3FGL J1442.8+1200}\\
{\footnotesize RX\,J0648.7+1516} & {\footnotesize 0.179} & {\footnotesize HBL} & {\footnotesize 16.6 } & {\footnotesize 3FGL J0648.8+1516}\\
{\footnotesize RBS\,0413} & {\footnotesize 0.190} & {\footnotesize HBL} & {\footnotesize 17.3} & {\footnotesize 3FGL J0319.8+1847}\\
{\footnotesize 1ES\,0502+675} & {\footnotesize 0.341} & {\footnotesize HBL} & {\footnotesize 17.9} & {\footnotesize 3FGL J0508.0+6736}\\
{\footnotesize 3C\,66A} & {\footnotesize  $0.33 < z <  0.41$} & {\footnotesize IBL} & {\footnotesize 15.6$^{\dagger}$} & {\footnotesize 3FGL J0222.6+4301}\\
{\footnotesize PKS\,1424+240} & {\footnotesize $z > 0.604$} & {\footnotesize IBL} & {\footnotesize 15.0} & {\footnotesize 3FGL J1427.0+2347}\\
{\footnotesize RGB\,J2243+203} & {\footnotesize ?} & {\footnotesize HBL} & {\footnotesize 15.1} & {\footnotesize 3FGL J2243.9+2021}\\
\hline
\end{tabular}
\caption{The AGN discovered at VHE with VERITAS and their
  classification \cite{TEVCAT} and synchrotron peak frequency (from 2WHSP or \cite{Nieppola}$^{\dagger}$).
}
\label{blazar_table}
\end{table}

VERITAS has discovered VHE emission from the 12 blazars shown in Table~\ref{blazar_table}. Details on two of the most
recent discoveries (S3\,1227+25 and RGB
J2243+203), whose VERITAS observations were triggered
by flares observed in  {\it Fermi}-LAT data, can be found in
\cite{Benbow_ICRC2015}.  The VHE discovery of
{\bf 1ES\,1440+122}  by VERITAS (initially reported in 2010; ATel \#2786) was
presented in detail at the conference.  This
{\it Fermi}-LAT-detected HBL is at a redshift of $z = 0.163$, and has
also been classified as an IBL  suggesting it is probably a
borderline case.  It was identified as a likely VHE emitter
on the basis of its SED \cite{Luigi_AGN}, and its MeV-GeV properties
($\Gamma_{3FGL} \sim1.8$ and $\Gamma_{2FHL} \sim 2.8$)
also motivate its VHE observation.   From 2008-2010, VERITAS acquired
$\sim$53 h of good-weather observations of 1ES\,1440+122
\cite{VERITAS_1440}.   An excess of 166 $\gamma$-rays
($\sim$5.5$\sigma$) was detected  from the direction of the blazar.  The observed flux above 200 GeV is
steady at $\sim$1.2\% Crab and the measured photon index is $\Gamma =
3.1 \pm 0.4$.  The SED generated from contemporaneous VERITAS, 
{\it Fermi}-LAT, and {\it Swift} UVOT / XRT data can be fit by SSC, SSC+EC,
and lepto-hadronic models.  The SSC+EC model is notably close to
equipartition while the other two are not, which is similar to the
behavior inferred by VERITAS from other VHE IBLs.  

{\bf Upper limits} were
recently published from  $\sim$570 hours of good-quality VERITAS observations (2007-2012) of 93 blazars \cite{VERITAS_BlazarUL_I}.  
While none of these individual sources were detected, a 4.6$\sigma$ (pre-trials)
excess is seen if one stacks the observations of all 36 relatively nearby ($z<0.6$) HBLs in the sample, i.e. the dominant population of extra-galactic
VHE sources.  No significant excess is seen (0.6$\sigma$) by stacking the remaining targets.  
Figure~\ref{AGN_mix_plot1} shows the distribution of integral flux upper
limits derived from VERITAS observations of each of these blazars, as
well as for the 21 2FGL sources serendipitously located in the
3.5$^{\circ}$ field of view of VERITAS. This typical limit from this
sample is $\sim$2\% Crab, and often the most sensitive produced at VHE.
It is also notable that this sample is larger than the combined sample
for which limits have been published by all third-generation VHE
instruments.  Despite the scale of this initial study, another
publication is in preparation with limits on $\sim$75 targets from
2012-2016 VERITAS discovery observations.  A comprehensive
VHE discovery study that should include observations of the
most-compelling blazars remaining from the 3FGL, 2FHL
and 2WHSP catalogs (as well as any ToO observations) is expected to be
completed by 2019.

The VERITAS collaboration makes extensive efforts to understand the
acceleration mechanisms within blazars.  To this end, every VERITAS
blazar detection has contemporaneous MWL data to enable source
modeling.  In general, the one-zone SSC model describes these
data well, even during flares.  However, SSC plus EC models
are favored for many VERITAS IBL detections on the basis of
equipartition arguments. In addition, for some of the extreme HBLs (i.e. those with the highest frequency
synchrotron peaks) detected by VERITAS a lepto-hadronic scenario is
weakly favored.  However, the statistics are generally low and no
strong claims can be made.
Although the SSC models have long been successful in describing the
SEDs of VHE blazars, the recent VERITAS campaign on 
1ES\,0229+200 shows that significant progress can still be made with
these models  \cite{1ES0229_paper}. This extreme HBL was observed by VERITAS for 54.3 h from 2009-2012
as part of a MWL monitoring campaign including {\it Swift}, RXTE and
{\it Fermi}-LAT.   Rather than presenting a single, degenerate solution for the modeling of the resulting MWL data, the entire range of
allowed SSC parameters were calculated.  Although 
the detection was not particularly strong in any given waveband, in many cases the SSC parameters were
constrained to a factor of $\sim$2 showing the general promise for
VERITAS blazar campaigns.

\section{Long-term AGN Observing Strategy}
\begin{figure}[t]
\centering
   \centerline{ {\includegraphics[width=4.0in]{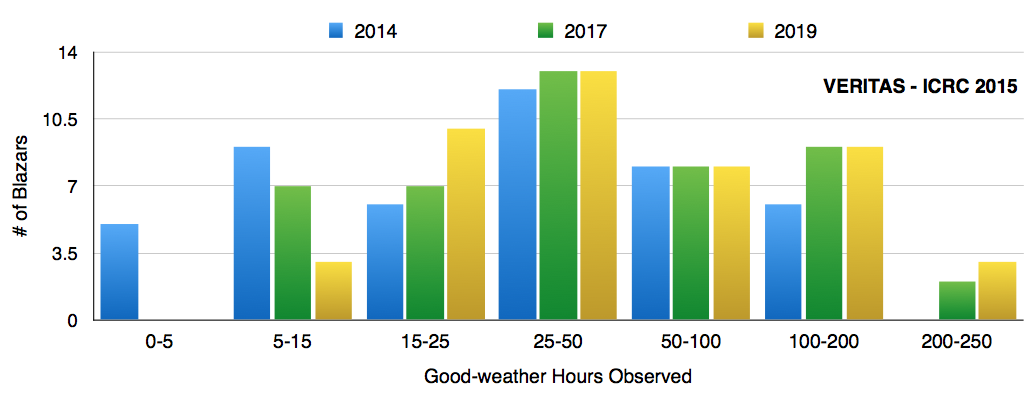} }}
 \caption{The histogram of the actual, or minimally expected, VERITAS exposure
    for each Northern Hemisphere blazar in July of 2014, 2017 and
    2019.  No observations beyond the monitoring programs (e.g. ToO
    data) are assumed.  Three blazars not known to be VHE
    emitters in July 2014 are included.}
  \label{Min_exposure}
 \end{figure}

The VERITAS collaboration began a five-year AGN observation program in Fall 2014.  
Overall, the goal of this program is to provide regular sampling of the light
curves of all $\sim$50 Northern-Hemisphere VHE AGN for a five year period,
and in $\sim$15 cases for a $\sim$10-year period, all with intense MWL coverage.
Figure~\ref{Min_exposure} shows a histogram of the minimum total
VERITAS exposure for each of the Northern-Hemisphere VHE blazars
that will eventually be acquired under this plan, as well as where the 
exposures stood at the start of the program.  As can be seen, every VHE
blazar visible to VERITAS will have reasonably deep coverage by 2019,
which should enable modeling efforts similar to those performed for 1ES\,0229+200.

The following describes the general AGN program which is reviewed and
receives minor updates every summer.  For blazars, it effectively has two key
components: deep observations of 10 key targets via intense, regular
monitoring and a {\it Snapshot Program}.  The deep observations are
broken into three core programs:  One program focuses on {\it EBL and Cosmic-ray
Line of Sight Measurements} via observations of five moderately
distant, hard VHE spectrum blazars.  Another core program focuses on
{\it Understanding MWL Variations} via observations of two IBLs that 
are both highly variable at all wavelengths,
and detected by VERITAS during their low states \cite{VERITAS_IBL}.  Another core program will focus on
generating regular observations of three {\it Iconic Objects}.
For the {\it Snapshot Program}, every week VERITAS will briefly
observe any of the other visible objects remaining in the Northern VHE
blazar catalog ($\sim$40 targets), to detect flaring events.  
While the duration of each target's snapshot varies
(typically 15 min, up to 1 h), the minimal detection sensitivity of any snapshot is
10\% Crab flux.  An automatic, real-time analysis pipeline and
a comprehensive decision tree for ToO-triggering ensures
that instantaneous follow-up of any flare at least five times the
base-line flux (minimally 10\% Crab) occurs. 
For about one-third of the Northern VHE
blazar catalog, the aforementioned monitoring
observations are taken simultaneously with {\it Swift} UVOT / XRT.
In addition, for each of the Northern VHE blazars intense
BVri coverage is arranged with the FLWO 48-inch optical telescope.

The blazar monitoring program (deep observations and {\it Snapshot Program})
has $\sim$210 h allocated.  The radio-galaxy component
of the AGN program consists of monitoring of the known VHE
emitters NGC\,1275 (snapshots) and  M\,87 (deep exposure as part of a global MWL campaign),
and has $\sim$20 h / yr allocated.  In addition, an $\sim$80 h annual allocation is
pre-approved for ToO follow-up of flaring AGN events meeting any of a number of MWL /
VHE triggers; the pre-approval exists to reduce logistical hurdles for
triggering, including instantaneous ones. This allocation matches the
average quantity of blazar ToO data taken each season and additional ToO time is possible.  
Additional observations ($\sim$100 h / year during dark time and $\sim$175
h / year during bright-moon time) are envisioned for {\it Other AGN Projects}
that are proposed annually.  

\section{Conclusions}
AGN observations remain a major component ($\sim$50\%) of the 
scientific program of VERITAS.  Thirty BL Lac objects, two FSRQs
and two FR-I radio galaxies are detected with the observatory, and
constraining upper limits are published from VERITAS observations of
$\sim$100 other AGN.  VERITAS is running well and its site operations 
are funded through 2019.  A strategy guiding
the VERITAS AGN program through 2019 is correspondingly organized, and is heavily 
focused on regular VHE and MWL monitoring of all known VHE AGN in the Northern Hemisphere, and
intense ToO follow-up of interesting flaring events.  The
collaboration continues to search for new VHE blazars, largely with ToO observations
during ``dark time'' and non-ToO bright-moon observations.  Given the
significant commitment of the VERITAS collaboration to AGN research,
there  should be many exciting results still to come.

% Acknowledgement
\section{ACKNOWLEDGMENTS}
This research is supported by grants from the U.S. Department of
Energy Office of Science, the U.S. National Science Foundation and the
Smithsonian Institution, and by NSERC in Canada. We acknowledge the
excellent work of the technical support staff at the Fred Lawrence
Whipple Observatory and at the collaborating institutions in the
construction and operation of the instrument.  The VERITAS
Collaboration is grateful to Trevor Weekes for his seminal
contributions and leadership in the field of VHE gamma-ray
astrophysics, which made this study possible.

% References

\nocite{*}
\bibliographystyle{aipnum-cp}%
\bibliography{Benbow_HDGS_AGN}%

\end{document}